\documentclass[paper]{frontiersSCNS} 

\usepackage{url,hyperref,lineno,microtype,subcaption}
\usepackage[onehalfspacing]{setspace}
\usepackage{tabu}
\usepackage{amssymb}
\usepackage{amsmath}
\usepackage{graphicx}

\def\keyFont{\fontsize{8}{11}\helveticabold }
\def\firstAuthorLast{Crutcher \& Kemball} 
\def\Authors{Richard M. Crutcher\,$^{1,*}$ and Athol J. Kemball\,$^{1}$}
\def\Address{$^{1}$Department of Astronomy, University of Illinois, Urbana, IL, USA}
\def\corrAuthor{1002 W. Green St., Urbana, IL 61801 USA}

\def\corrEmail{crutcher@illinois.edu}

\begin{document}
\onecolumn
\firstpage{1}

\title[Zeeman Observations]{Review of Zeeman Effect Observations of Regions of Star Formation} 

\author[\firstAuthorLast ]{\Authors} 
\address{} 
\correspondance{} 

\extraAuth{}

\maketitle

\begin{abstract}

\section{}
The Zeeman effect is the only observational technique available to measure directly the strength of magnetic fields in regions of star formation. This chapter reviews the physics of the Zeeman effect and its practical use in both extended gas and in masers. We discuss observational results for the five species for which the Zeeman effect has been detected in the interstellar medium -- H~I, OH, and CN in extended gas and OH,  CH$_3$OH, and H$_2$O in masers. These species cover a wide range in density, from $\sim10$ cm$^{-3}$ to $\sim10^{10}$ cm $^{-3}$, which allows magnetic fields to be measured over the full range of cloud densities. However, there are significant limitations, including that only the line-of-sight component of the magnetic field strength can usually be measured and that there are often significant uncertainties about the physical conditions being sampled, particularly for masers. We discuss statistical methods to partially overcome these limitations. The results of Zeeman observations are that the mass to magnetic flux ratio, which measures the relative importance of gravity to magnetic support, is subcritical (gravity dominates magnetic support) at lower densities but supercritical for $N_H \gtrsim 10^{22}$ cm$^{-2}$. Above $n_H\sim 300$ cm$^{-3}$, which is roughly the density at which clouds typically become self-gravitating, the strength of magnetic fields increases approximately as $B \propto n^{2/3}$, which suggest that magnetic fields do not provide significant support at high densities. This is consistent with high-density clouds being supercritical. However, magnetic fields have a large range in strengths at any given density, so the role of magnetic fields should differ significantly from one cloud to another. And for maser regions the dependence of field strength on density may have a slightly lower slope.  Turbulent reconnection theory seems to best match the Zeeman observational results. 
\tiny
 \keyFont{ \section{Keywords:} Zeeman effect, magnetic fields, molecular clouds, mass/flux ratio, star formation, masers} 
\end{abstract}

\newpage

\section{INTRODUCTION}\label{introduction}

What governs or regulates star formation has been a crucial question in astrophysics for many decades. The two extreme positions are: (1) that magnetic fields support clouds against gravitational collapse and that star formation can occur only when magnetic support is removed, through a process such as ambipolar diffusion, e.g. \cite{1999ASIC..540..305M}
or (2) that interstellar turbulence governs the formation of self-gravitating clouds that once formed can collapse and form stars, e.g. \cite{2004RvMP...76..125M}. 
Theory alone cannot answer this question; required are observations of magnetic fields. 

There are several observational techniques for the study of magnetic fields in the interstellar medium, with the two most prominent being (1) linear polarization of continuum radiation emitted or absorbed by dust grains aligned by magnetic fields and (2) the Zeeman effect that produces frequency-shifted polarized spectral lines. This chapter is concerned with the Zeeman effect. 

The Zeeman effect was discovered by Dutch physicist Pieter Zeeman in 1896 in a laboratory experiment. In his discovery paper Zeeman suggested that the effect he had discovered could be important in measuring magnetic fields in astrophysics. The Zeeman effect was first applied in astrophysics in 1908 by George Ellery \cite{1908ApJ....28..315H}, who measured magnetic fields in sun spots. Although interstellar magnetic fields were first detected by \cite{1949Sci...109..165H} by observing linear polarization of starlight passing through the intervening interstellar medium, the first detection of the Zeeman effect in the interstellar medium came only after almost another 20 years. The first indication of an interstellar Zeeman effect came from observations of polarization in OH masers by \cite{1965Natur.208..440W}; although they suggested that the polarization might be due to the Zeeman effect, that interpretation was not certain because the standard pattern of the classical Zeeman effect was not seen in the polarized maser emission. Following intense observational work by several workers, \cite{1968PhRvL..21..775V} first detected the Zeeman effect in the extended interstellar medium in the 21-cm hyperfine line of H~I. It was another 15 years before Zeeman splitting in extended molecular gas was detected, in OH by \cite{1983A&A...125L..23C}, and yet another 15 years before detection in the third (and so far last) species, CN, by \cite{1996ApJ...456..217C, 1999ApJ...514L.121C}. Interstellar maser observations of the Zeeman effect have been extended from OH to additional species: H$_2$O \citep{1989LNP...331...93F} and CH$_3$OH \citep{vlemmings_new_2008}. 

The Zeeman effect is used to study magnetic fields in the diffuse and dense (molecular) interstellar medium. In this chapter we review Zeeman observations in the interstellar medium and discuss how observations of the Zeeman effect can test models of star formation, the present state of such tests, and possible future developments. 
\cite{2012ARA&A..50...29C} previously reviewed observations (by all techniques) of magnetic fields in molecular clouds. This article is specific to the Zeeman effect; it expands discussion of the effect itself, summarizes the discussion in the above review, adds the (very limited) new Zeeman data that have become available, and discusses some more recent controversies about the astrophysical interpretation of the observational results.

\section{THE ZEEMAN EFFECT}\label{zeeman_physics}

Immediately after Zeeman's discovery, Hendrik Lorentz explained the Zeeman effect in terms of classical physics -- an electrical charge moving in a circular orbit in a magnetic field. The predicted frequencies for the Zeeman-split lines are then: 
\begin{equation}
\nu = \nu_0 \pm \frac{eB}{4\pi m_ec},
\label{classicalZeeman}
\end{equation}
where $\nu_0$ is the unshifted line frequency, $e$ is the charge of an electron, $B$ is the magnetic field strength, $m_e$ is the mass of an electron, and $c$ is the speed of light.

The above might suggest that in the presence of a magnetic field, an atom would emit two spectral lines at the frequencies given by eq. \ref{classicalZeeman}. However, only the vector component of the electron acceleration perpendicular to the line of sight will produce electromagnetic radiation along the line of sight. For atoms with electron orbital planes perpendicular to the line of sight, as the electrons accelerate in their circular orbits they will emit circularly polarized radiation along the magnetic field direction. Since electrons are negatively charged, right circularly polarized radiation has the higher frequency and left circularly polarized radiation the lower frequency. On the other hand, if one observes perpendicular to the magnetic field vector {\bf B} with the electron orbits perpendicular to {\bf B}, then only the acceleration of the electron perpendicular to {\bf B} will emit radiation along the line of sight, and this radiation will be linearly polarized perpendicular to {\bf B}. In practice one could not observe a single atom but an ensemble of atoms with electron orbital planes randomly distributed with respect to {\bf B}. In the case of observations along the magnetic field direction, the projection of these orbits onto the plane perpendicular to {\bf B} would produce the same two circularly polarized Zeeman components as described above. For observations perpendicular to {\bf B}, however, a frequency of radiation different from the two Zeeman components is introduced. Electron orbits parallel to {\bf B} will produce linearly polarized radiation parallel to {\bf B} but unshifted in frequency since the electron acceleration that produces the radiation is parallel to {\bf B}, hence there is no additional force produced by the magnetic field.

However, with higher spectral resolution it was soon found that the triplet of lines predicted by classical physics was actually a more complicated pattern of more than three lines. This ``anomalous'' Zeeman effect is not explicable by classical physics, for it is only the angular momentum of the electron in its orbit that produces the classical Zeeman effect. Electron spin means that it is the total angular momentum, both orbital and spin angular momenta, that produces the Zeeman effect. The various coupling modes of the two angular momenta produce the more complicated anomalous effect with more than the three classical line components when the net spin of the electrons is an odd half integer. Therefore atomic hydrogen is in general a case of anomalous Zeeman splitting. However, since the 21-cm line of H~I arises from the two hyperfine energy levels in the ground S$_{1/2}$ state with $m_J = \pm 1/2$, the Zeeman pattern is the classical three-line one. The ``anomalous'' Zeeman effect is in fact THE Zeeman effect; the ``normal'' Zeeman effect with a triplet of lines is simply a subset of the complete situation. With a single electron moving in a magnetic field, the prediction is the same as the classical one:
\begin{equation}
\nu = \nu_0 \pm \frac{\mu_BB}{h}, 
\label{quantumZeeman}
\end{equation}
where the Bohr magneton $\mu_B \equiv e h/4\pi m_ec = 9.2732 \times 10^{-21}$  erg/G, which means that the frequency shift due to the Zeeman effect for the 21-cm H~I line is 1.4 Hz/$\mu$G. 

The atoms and molecules that are most useful in studying magnetic fields in the interstellar medium are hydrogenic, with an odd number of electrons. The unpaired electron in hydrogenic systems will lead to large Zeeman splittings, approximately equal to the Bohr magneton, and hence to detectable Zeeman splittings at the relatively low ($\sim 10$ $\mu$G in H~I, e.g. \citealt{2005ApJ...624..773H}) magnetic field strengths in the interstellar medium. Even for systems with all electrons paired, there may still be a Zeeman effect, but this time due to the nuclear spin rather than the electron spin. However, the Zeeman splitting is now approximately equal to the nuclear magneton,
$\mu_N = eh/4\pi m_pc = \mu_B/1836$. Only for very strong magnetic fields and very strong spectral lines is the Zeeman effect detectable in the interstellar medium for this case; examples are the CH$_3$OH and H$_2$O masers. 

Figure~\ref{ZeemanEnergy&Pol} (upper) illustrates the above discussion of the Zeeman effect. The lower part of figure~\ref{ZeemanEnergy&Pol} shows the polarizations of the frequency unshifted ($\pi$) and shifted ($\sigma$) Zeeman components, with their polarization states depending on the viewing angle. 

\section{OBSERVING THE ZEEMAN EFFECT}\label{zeeman_observing}

Figure~\ref{ZeemanProfiles} illustrates what will be observed for the classical Zeeman effect.  Figure~\ref{ZeemanProfiles}(a) shows a Gaussian Stokes I spectral line with unit intensity and unit sigma-width $\Delta\nu_\sigma$ (i.e., FWHM = 2.355). Figure~\ref{ZeemanProfiles}(b) shows the three Zeeman components when the Zeeman splitting is sufficiently large that the Zeeman components are cleanly separated, with the total intensity of each component being in the ratio 1:2:1. Figure~\ref{ZeemanProfiles}(c) shows what would be observed with a instrument sensitive to circular polarization with a 1\% Zeeman splitting. Figure~\ref{ZeemanProfiles}(d) shows what would be observed with a instrument sensitive to linearly polarization perpendicular to the magnetic field in the plane of the sky with a large Zeeman splitting, such as might occur in mainline OH masers. In these cases, one would observe the Zeeman components with their full separations $\Delta\nu_z$ and be able to infer the full magnetic field strength. However, if the Zeeman splitting were only 1\% of the line width, Figure~\ref{ZeemanProfiles}(e) shows the Stokes Q and/or U spectra that would be observed. Note the very small amplitude of the signal. Finally, figure~\ref{ZeemanProfiles}(f) complements figure~\ref{ZeemanProfiles}(d) and shows the spectrum that would be observed in linear polarization observing parallel to a field in the plane of the sky.

The strength of the $\pi$ component is proportional to the strength of the magnetic field in the plane of the sky parallel to the magnetic vector {\bf B}. The $\sigma$ components are generally elliptically polarized, since the magnetic field will in general be at an angle to the line of sight. The elliptical polarization is a combination of linear polarization perpendicular to {\bf B} in the plane of the sky and circular polarization proportional to the strength of the magnetic field along the line of sight. The sense of the circular polarization of the two $\sigma$ components reverses depending on whether the line-of-sight component $B_{LOS}$ is toward or away from us. From $\Delta\nu_z$, the degree of elliptical polarization of the $\sigma$ components, and the relative amplitudes of the $\sigma$ and $\pi$ components, it is possible in principle to infer full information about {\bf B} \citep{1993ApJ...407..175C}. 

However, in the extended (non-masing) interstellar medium, the Zeeman splitting is generally much smaller than the line width, and it is possible to infer only the amplitude and direction of the line-of-sight component $B_{LOS}$ of {\bf B}.  We can see why this is if we consider a magnetic field with components both along the line of sight and in the plane of the sky. If an instrument is sensitive only to (for example) left-circular polarization, it could detect (say) 100\% of the $\sigma-$ (or equivalently red-shifted $\sigma_r$) Zeeman component and 0\% of the $\sigma+$ (or equivalently blue-shifted $\sigma_b$) component. But the linearly polarized $\sigma-$, $\pi$, and $\sigma+$ Zeeman components would also be detected by this instrument at a fraction of their full intensity (depending on $\theta$, the angle between the line of sight and {\bf B}). This would increase the detected strength of the $\sigma+$ Zeeman component, but would shift the centroid of the observed left-circularly polarized spectrum toward the unshifted line frequency $\nu_0$. Hence, the observed Zeeman frequency shift would be less than $\Delta\nu_z$. When the right circularly polarized spectrum was observed, a similar shift toward the central unshifted frequency would occur. The result would be an ``observed'' Zeeman splitting less than $\Delta\nu_z$, by an amount $cos\theta$. Hence, observation of the Stokes parameter V spectrum yields only $B\cos\theta = B_{LOS}$, the line-of-sight component of {\bf B}. As illustrated in figure~\ref{ZeemanProfiles}(c), when the Zeeman splitting is 1/100th the width of the spectral line ($\Delta\nu_z = 0.01 \Delta\nu_\sigma$), the shape of the Stokes V spectrum is that of the first derivative with respect to frequency of the Stokes I spectrum and the total amplitude of the Stokes V spectrum is about 1/100 that of the Stokes I spectrum. 

In principle, information about the field in the plane of the sky (POS) would come from the Stokes Q and U spectra, with strengths proportional to $(\Delta \nu_z / \Delta \nu_\sigma)^2 \times B_{POS}$. The reason for this second-order dependence of the strengths of the Stokes Q and U spectra on the strength of the magnetic field comes from the fact that there is not just the Zeeman-split linearly-polarized $\sigma$ components, but also the linearly polarized and non-shifted $\pi$ Zeeman component. Figure~\ref{ZeemanProfiles}(e) illustrates this for a Zeeman splitting of 1/100th the width of the spectral line. The magnitude of the Stokes Q or U spectrum (depending on the orientation of the magnetic field on the plane of the sky) is much too small to be detected as a practical matter. 

To date the Zeeman effect has been detected unambiguously in non-masing interstellar gas only in H~I, OH, and CN lines and in maser lines of OH, CH$_3$OH, and H$_2$O. The Zeeman splitting factor $Z = 2\Delta\nu_z$, which is the separation between the two $\sigma$ components or twice the Zeeman frequency shift, is specific to the spectral transition. The $Z$s  for transitions with interstellar Zeeman detections are given in table~\ref{Zs} (maser lines of CH$_3$OH and H$_2$O are blends of several $\Delta$F transitions, and $Z$ factors are somewhat uncertain). In this table R.I. is the relative intensity and Z is the Zeeman splitting factor of each line, so R.I. $\times$ Z is the relative sensitivity to $B_{LOS}$. Other promising species are C$_2$H, SO, C$_2$S, and CH. Unfortunately, most of the common interstellar molecules have all their electrons paired (non-paramagnetic species) and therefore do not have strong $Z$ factors. Because $Z$ does not depend on spectral-line frequency, sensitivity to the Zeeman effect decreases with increasing spectral-line frequency, so cm-wavelength transitions like those of H~I and OH are sensitive to much lower field strengths than those of mm-wavelength transitions like those of CN. 

\begin{table}
\caption{Zeeman Splitting Factors Z }
\vspace{8pt}
\begin{tabular}{ccccccc}  
\underline{Species} & \underline{Transition} & \underline{$\nu$ (GHz)} & \underline{R.I.} & \underline{Z (Hz/$\mu$G)} & \underline{R.I. $\times$ Z} & \underline{Ref} \\ [8pt]
H~I & $F = 1-2$ & 1.420406 & 1 & 2.80 & - & 1\\  [8pt]
CH & $J=^3\!/\!_2,F=2-2$ & 0.701677 & 5 & 1.96 & 16.4 & 1\\  [2pt]
  & $J=^3\!/\!_2,F=1-1 $ & 0.7724788 & 9 & 3.27 & 17.6 & 1\\  [8pt]
OH & $J=^3\!/\!_2,F=1-1$ & 1.665402 & 5 & 3.27 & 16.4 & 1\\  [2pt]
  & $J=^3\!/\!_2,F=2-2 $ & 1.667359 & 9 & 1.96 & 17.6 & 1\\  [2pt]
  & $J=^3\!/\!_2,F=2-1 $ & 1.720530 & 1 & 1.31 & 1.31 & 1\\  [8pt]
  & $J=^5\!/\!_2,F=2-2 $ & 6.030747 & 7 & 1.58 & 11.1 & 1\\  [2pt]
 & $J=^5\!/\!_2,F=3-3 $ & 6.035092 & 10 & 1.13 & 11.3 & 1\\  [8pt]
  & $J=^7\!/\!_2,F=3-3 $ & 13.434637 & 27 & 1.03 & 28 & 2\\  [2pt]
 & $J=^7\!/\!_2,F=4-4 $ & 13.441417 & 35 & 0.80 & 28 & 2\\  [8pt]
 CH$_3$OH & $J_N = 5_1-6_0$ & 6.668519 & 1 & -0.00114 & - &3\\  [8pt]
CCS & $J_N = 1_0-0_1$ & 11.119446 & 1 & 0.81 & 0.81 &4\\  [2pt]
 & $J_N = 2_1-0_1$ & 22.344033 & 1 & 0.77 & 0.77 &4\\  [2pt]
 & $J_N = 3_2-2_1$ & 33.751374 & 1 & 0.70 & 0.70 &4\\  [2pt]
 & $J_N = 4_3-3_2$ & 45.379033 & 1 & 0.63 & 0.63 & 4\\  [8pt]
H$_2$O & $F=6_{16}-5_{23}$ & 22.23508 & 1 & 0.003 & - &5\\  [8pt]
SO & $J_N = 1_0-0_1$ & 30.001630 & 1 & 1.74 & 1.74 &4\\  [2pt]
 & $J_N = 1_2-0_1$ & 62.931731 & 1 & 0.93 & 0.93 & 4\\  [2pt]
 & $J_N = 1_1-2_2$ & 86.094 & 1 & 1.38 & 1.38 &6\\  [2pt]
 & $J_N = 3_2-2_1$ & 99.299875 & 1 & 1.04 & 1.04 &4\\  [2pt]
 & $J_N = 4_3-3_2$ & 138.178548 & 1 & 0.80 & 0.80 &4\\  [2pt]
 & $J_N = 3_2-4_3$ & 158.972 & 1 & 0.81 & 0.81 & 6\\  [8pt]
CCH & $N=1-0,J=^3\!/\!_2-^1\!/\!_2,F=2-1$ & 87.31723 & 42 & 0.70 & 29 &7\\  [2pt]
  & $N=1-0,J=^3\!/\!_2-^1\!/\!_2,F=2-1$ & 87.32892 & 21 & 2.3 & 48 &7\\  [2pt]
  & $N=1-0,J=^3\!/\!_2-^1\!/\!_2,F=2-1$ & 87.40234 & 21 & 0.93 & 20 & 7\\  [8pt]
CN & $J=^1\!/\!_2-^1\!/\!_2, F=\,^1\!/\!_2-^3\!/\!_2$ & 113.1442 & 8 & 2.18 & 17.4 &8\\  [2pt]
 & $J=^1\!/\!_2-^1\!/\!_2, F=\,^3\!/\!_2-^1\!/\!_2$ & 113.1705 & 8 & -0.31 & 2.5 &8\\  [2pt]
 & $J=^1\!/\!_2-^1\!/\!_2, F=\,^3\!/\!_2-^3\!/\!_2$ & 113.1913 & 10 & 0.62 & 6.2 &8\\ [2pt]
 & $J=^3\!/\!_2-^1\!/\!_2, F=\,^3\!/\!_2-^1\!/\!_2$ & 113.4881 & 10 & 2.18 & 21.8 &8\\  [2pt]
 & $J=^3\!/\!_2-^1\!/\!_2, F=\,^5\!/\!_2-^3\!/\!_2$ & 113.4910 & 27 & 0.56 & 15.1 &8\\  [2pt]
 & $J=^3\!/\!_2-^1\!/\!_2, F=\,^1\!/\!_2-^1\!/\!_2$ & 113.4996 & 8 & 0.62 & 5.0 &8\\ [2pt]
 & $J=^3\!/\!_2-^1\!/\!_2, F=\,^3\!/\!_2-^3\!/\!_2$ & 113.5089 & 8 & 1.62 & 13.0 & 8\\  [12pt]
\end{tabular}
\label{Zs}
References: 1. \cite{1993prpl.conf..279H}, 2. \cite{2001A&A...371..274U}, 3. \cite{lankhaar_characterization_2018}, 4. \cite{Shinnaga_2000}, 5. \cite{1992ApJ...384..185N}, 6. \cite{2017A&A...605A..20C}, 7. \cite{1998A&A...335.1025B}, 8. \cite{1996ApJ...456..217C}
\end{table}

Zeeman radio spectral-line observations are generally not reported as right and left circularly polarized spectra (RCP and LCP), but as Stokes parameter spectra I = RCP + LCP and V = RCP - LCP (for technical reasons telescope software sometimes returns I and V as 1/2 the above). As noted above, for the limit of Zeeman splitting much smaller than the spectral-line width, the ``theoretical'' Stokes V spectrum is $V \approx dI/d\nu \times Z \times B\cos\theta$. The analysis technique followed is to calculate the derivative of the
I spectrum by numerically differentiating the observed I spectrum, and to fit (usually by least squares) this to the observed V spectrum. Because there may be a gain difference ``g'' between the left and right circular polarization due to instrumental effects, generally one also includes a term to fit for this. So the general equation fitted to the observed Stokes V spectrum is
\begin{equation}
V \approx dI/d\nu \times Z \times B\cos\theta + g \times I.
\end{equation}
The mean error in $B\cos\theta$ is also given by the least-squares fitting process.

Very small instrumental polarization effects can be very
important for Zeeman work; hence, observers must be very careful that instrumental polarization effects are not mistaken for the Zeeman effect. The most significant instrumental effect in single-dish Zeeman observations is the phenomenon of
``beam squint'', for which the left and right circularly polarized
beams of the telescope point in different directions. Beam squint is important
for Zeeman work because the combination of beam squint
and a velocity gradient in a cloud will produce a V spectrum
identical to the one expected for the Zeeman effect. \citet{2004ApJS..151..271H} have extensively described the possible instrumental effects and techniques for mitigating these effects when performing Zeeman observations. The technical challenges of high-accuracy Stokes V interferometric observations and analysis are described by \citet{sault_1990} and \citet{2011A&A...533A..26K}.

When only one component of {\bf B} is measurable, the Zeeman effect gives directly only a lower limit to the total magnetic field strength. However, a statistical study of a large number of clouds can yield information about total field strengths, e.g., \cite{2005LNP...664..137H}. Most of the earlier statistical studies assumed that the measured $B_{LOS}$ in a set of clouds were uniformly distributed between $-B_0$ and $B_0$, where $B_0$ is the total field strength that is assumed to be the same in all lines of sight observed. Then the median and mean of the set of measured $\vert B_{LOS}\vert = \frac{1}{2}B_0$. Hence, one simply determines the mean of $\vert B_{LOS}\vert$ to infer $B_0$. However, the approaches that deal only with mean or median values ignore the possibly large variation in total field strength ($B_{TOT}$) in the sample. A more sophisticated approach is to measure the probability distribution function (PDF) of $B_{LOS}$ over a sample of clouds and to infer the PDF of the total magnitude of the magnetic field strength. An application of this approach will be discussed in detail below.

\section{Zeeman Observational Results - Extended Gas}

\subsection{H~I, OH, and CN Zeeman Observations}

The Zeeman effect in the ISM was first detected -- after multiple attempts -- in absorption lines of H~I toward the Cassiopeia A supernovae remnant \citep{1968PhRvL..21..775V}. Over the next five years only three more detections were made, toward Orion A, M 17, and Taurus A. Three of these were in the H~I associated with molecular clouds, while the Taurus A line is not. 

\cite{1982ApJ...260L..19T,1982ApJ...252..179T} and \cite{1982ApJ...260L..23H} achieved the first H~I Zeeman detections beyond Verschuur's original four souces, 14 years after that first detection.  Further emission-line H~I Zeeman observations and maps were toward the dark cloud filament L204 \citep{1988ApJ...324..321H}, H~I filaments associated with supernova or super-bubble shells \citep{1989ApJ...336..808H}, the Ophiuchus dark cloud \citep{1994ApJ...424..208G}, and four dense H~I clouds \citep{1995ApJ...442..177M}. \cite{1997ApJS..111..245H} mapped H~I Zeeman toward 217 positions in the Orion-Eridanus region and carried out an extensive analysis. Finally, \cite{2004ApJS..151..271H} carried out a large H~I Zeeman survey in absorption lines toward continuum sources.

The first detection was of OH absorption toward the NGC 2024 molecular cloud \citep{1983A&A...125L..23C}. The OH Zeeman effect was later mapped with the VLA (e.g., figure \ref{NGC2024}) toward several molecular clouds. 

\cite{1993ApJ...407..175C} carried out a survey of OH Zeeman toward dark clouds, achieving mostly upper limits. \cite{2001ApJ...554..916B} extended attempts to detect the OH Zeeman effect, and obtained one definite and one probable new detection out of the 23 molecular clouds observed. Then \cite{2008ApJ...680..457T} carried out a major survey toward dark clouds, with 9 detections out of 34 positions. 

\cite{1996ApJ...456..217C, 1999ApJ...514L.121C} detected the Zeeman effect in a second molecular species, CN. Finally, \cite{2008A&A...487..247F} extended the earlier work on CN Zeeman with a survey of dense molecular cores. The combined total was 14 positions observed and eight detections. 

Figures \ref{BvsN} and \ref{Bvsn} show results for the Zeeman observations of H~I, OH, and CN in extended gas.

\subsection{Interpretation of Zeeman Observations}

The three species (H~I, OH, and CN) with Zeeman detections in extended gas have resulted in measurements of $B_{LOS}$ that cover a large range of densities. H~I emission samples the cold neutral atomic medium over densities between 1-100 cm$^{-3}$. H~I in absorption toward molecular clouds can sample densities $\sim 10^2-10^4$ cm$^{-3}$; the ground-state 18-cm lines of OH sample roughly the same density range. Finally, the 3-mm emission lines of CN, which have a critical density $\sim 10^5$ cm$^{-3}$,  sample densities $\sim 10^5-10^6$ cm$^{-3}$. 

The astrophysical significance of Zeeman results requires determination of $N_H$ and/or $n_H$ in the regions where magnetic field strengths have been measured. For H~I in absorption $N_H$ may be determined by also observing the line in emission off the continuum source so that the spin temperature and optical depth can be inferred, e.g., \cite{2003ApJ...586.1067H}. The associated $n_H$ may then be estimated from the mean interstellar pressure in the cold neutral diffuse medium and the spin temperature, e.g., \cite{2010ApJ...725..466C}. Since the OH line optical depths are generally small, $N_{OH}$ can be estimated from the observed line strengths,  e.g., \cite{1979ApJ...234..881C}. To obtain $N_H$ one then uses the [OH/H] ratio determined by \cite{1979ApJ...234..881C}. To obtain $n_H$ for the regions in which OH is found one divides $N_H$ by the mean diameter of the OH region. For CN \citep{2008A&A...487..247F},  the methods are similar to those for OH. The CN hyperfine-line ratios imply that the lines are optically thin, so $N(CN)$ may be calculated from observed line strengths. $N_H$ then comes from [CN/H] based on studies by \cite{1975ApJ...198...71T} and \cite{2003A&A...412..157J}. The $n_H$ in the CN emitting regions must be fairly close to the critical density of the transition, since the lines are observed to be much weaker than kinetic temperatures and optically thin (no line photon trapping). Unfortunately few excitation analyses of CN excitation have been carried out, but since CN and CS have similar critical densities and map similarly, $n_H$ in the CN regions can be assumed to be about the same as obtained from CS excitation analyses. Finally, a second, independent method for determining $n_H$ comes by dividing $N_H$ by the estimated thickness of clouds from the mean extent of the CN distribution on the sky. There are certainly significant uncertainties in the estimates of $n_H$ especially, particularly as applied to individual clouds, where estimates may be off by an order of magnitude. However, in statistical studies such as those described in this paper, more important is the ensemble uncertainty. \cite{2010ApJ...725..466C} found a statistical uncertainty of about a factor of two in $n_H$

Two important quantities than can be inferred from the Zeeman data are the mass to magnetic flux ratio $M/\Phi$ ($\propto N_H/B$) and $\kappa$ (in the $B \propto n_H^\kappa$ relation (see \cite{2012ARA&A..50...29C} for a detailed discussion). $M/\Phi$ is proportional to the ratio of gravity to magnetic pressure and informs whether magnetic fields are sufficiently strong to support clouds against gravitational contraction. A simple way to derive the expression for the critical $M/\Phi$A at which magnetic and gravitational energies are in equilibrium is to equate the virial terms: $3GM^2/5R = B^2R^3/3$. Since magnetic flux $\Phi = \pi R^2B$, the critical $M/\Phi$ is:
\begin{equation}
\left(\frac{M}{\Phi} \right)_{critical} = \frac{1}{3\pi}\sqrt\frac{5}{G}.
\end{equation}
The precise numerical value differs slightly for detailed models depending on geometry and density structure. A supercritical ratio means that magnetic pressure alone is insufficient to prevent gravitational collapse, while a subcritical ratio means collapse is prevented by magnetic pressure. The scaling of magnetic field strength with density is a prediction of many theoretical studies of the evolution of the interstellar medium and star formation.  Simple examples include (1) mass accumulation along field lines without change in magnetic field strength, for which $\kappa = 0$;  compression of mass perpendicular to the field with flux freezing, for which $\kappa = 1$; and spherical collapse with flux freezing and weak field strength, for which $\kappa = 2/3$ \citep{1966MNRAS.133..265M}. 

\subsubsection{B vs. N}

First, we discuss field strength versus column density. \cite{2001ApJ...554..916B} plotted this for their OH observations and discussed the implication. Figure~\ref{BvsN} shows $B_{LOS}$ versus $N_H$ with data from the compilation by \cite{1999ApJ...520..706C} and four later major Zeeman surveys of H~I, OH, and CN \citep{2004ApJS..151..271H, 2001ApJ...554..916B, 2008ApJ...680..457T, 2008A&A...487..247F}. The data are clearly separated into three ranges in $N_H$, corresponding to the tracers H~I, OH, and CN. The straight line is the critical $M/\Phi$ line.

An essential point in interpreting figure~\ref{BvsN} is that only one component of the total magnetic vector{ \bf B} is measured. Hence, all points are lower limits on what the total magnetic field strength would be. However, for $N_H \lesssim 10^{21}$ cm$^{-2}$, most of the points are above the critical line, showing that at low column densities the diffuse H~I and lower column density molecular gas is subcritical. In contrast, for $N_H \gtrsim 10^{22}$ cm$^{-2}$, all but one of the points are below the critical line. It is possible that some of these clouds are subcritical with the magnetic field close to the plane of the sky. However, that fact that all of the points are below the critical line suggests strongly that a transition occurs at $N_H \sim 10^{22}$ cm$^{-2}$ from subcritical to supercritical $M/\Phi$. Clouds with $N_H \gtrsim 10^{22}$ cm$^{-2}$ have a mean $M/\Phi$ that is supercritical by a factor of 2--3. The data strongly suggest that subcritical self-gravitating clouds are the exception and in fact none may exist. These self-gravitating clouds are the ones in the ambipolar diffusion model that should be subcritical at early stages of gravitational contraction. 

Figure~\ref{BvsN} might appear to support the ambipolar diffusion model of cloud evolution, in which initially subcritical clouds become supercritical by gravitational contraction of neutral matter through magnetic fields. However, the points with $N_H \lesssim 10^{21}$ cm$^{-2}$, are lower density H~I clouds. These cold H~I clouds are confined by pressure from the surrounding warm ISM and are not self-gravitating, so they could not gravitationally collapse as envisioned by the ambipolar diffusion model. \cite{2005ApJ...624..773H} found that the mean $B_{TOT}$ is approximately the same in the cold H~I medium and the warm neutral medium. Hence, the magnetic field strength does not systematically change during transitions of gas between the lower density warm and the higher density cold neutral medium. Possible explanations for this are that diffuse clouds form by flows along magnetic flux tubes or that they form preferentially from regions of lower magnetic field strength. Another process that could be important in keeping field strengths fairly constant is turbulent magnetic reconnection \citep{1999ApJ...511..193V}. 

$N_H$ in the range $10^{21-22}$ cm$^{-2}$ marks a clear transition between magnetic field strengths being statistically independent of $N_H$ and an increase in strength with column density. A similar transition is seen in figure~\ref{Bvsn} (discussed below) at $n_H \approx 300$ cm$^{-3}$. Assuming that these $N_H$ and $n_H$ correspond to the same clouds, the typical diameters of these clouds is 0.1-1 pc. These are roughly the parameters for an interstellar cloud to become self-gravitating. Gravitational contraction with flux freezing would then cause the magnetic field strength to increase with increasing $N_H$ and $n_H$. We also note that $N_H \approx 10^{22}$ cm$^{-2}$ is also roughly the column density where the orientation of magnetic fields in the plane of the sky as mapped with polarized dust emission changes (statistically) from parallel to perpendicular with respect to the elongated mass structures on the plane of the sky \citep{2016A&A...586A.138P}. 

Probably the main uncertainty in figure~\ref{BvsN} comes from the column densities. For H~I the $N_H$ are very well determined, since both the line optical depths and spin temperatures are directly measured. However, for OH and CN the $N_H$ come from determinations of $N_{OH}$ and $N_{CN}$ and studies of OH/H and CN/H, which introduce possible errors. A major issue is exactly what $N_H$ the OH and CN Zeeman results sample. On the basis of ambipolar diffusion models with time-dependent astrochemistry, \cite{2012ApJ...754....6T} argue that OH and CN are heavily depleted at higher densities due to chemistry and hence tend to sample the lower density outer layers of clouds rather than the cores, and that therefore the Zeeman results underestimate the magnetic field strengths in cores. If the true field strengths are higher at each $N_H$ than those plotted in figure~\ref{BvsN}, many of the points with $N_H > 10^{22}$ cm$^{-2}$ should be plotted at stronger field strengths. Such points would then lie above the critical $M/\Phi$ line, and would represent subcritical self-gravitating clouds. One issue with this conclusion is that ambipolar diffusion driven evolution is significantly slower than those for which the magnetic flux problem has been resolved by other physics such as turbulent reconnection \citep{1999ApJ...511..193V,2012ApJ...757..154L}; the chemical depletion at high densities may not have had sufficient time to be as significant as \cite{2012ApJ...754....6T} find. A more direct problem with their argument is that the interpretation of figure~\ref{BvsN} does not depend on OH and CN sampling the highest densities of molecular cores. The Zeeman effect estimates the magnetic field strength in the regions sampled by the Zeeman tracer (OH or CN), and the relevant $N_H$ and $n_H$ for estimating $M/\Phi$ are those sampled by the Zeeman species. There is no claim that either species samples the highest densities of cores. Ideally one might use a variety of Zeeman species that sample a range of densities in order to measure the change in $M/\Phi$ from envelope to core in  clouds. The fact that all Zeeman species do not trace the field in the cores, while true, does not invalidate our interpretation of figure~\ref{BvsN}.

\subsubsection{B vs. n}

The above discussion was limited by the fact that only the line-of-sight component of the vector {\bf B} is measured with the Zeeman effect. However, with a large number of Zeeman measurements, it is possible to infer statistical information about the total field strength. One can assume a PDF of the total field strength, $P(B_{TOT})$, and compute $P(B_{LOS})$, the PDF of the observable line-of-sight field strengths, assuming a random distribution of the $\theta$. Comparison between the two lets one infer the most probable (of those assumed) $P(B_{TOT})$. \cite{2005LNP...664..137H} attempted this for  H~I Zeeman data with a frequentist approach, but found that the observations did not allow a strong discrimination among possible PDFs for the total field strength. 

\cite{2010ApJ...725..466C} used a Bayesian approach, and expanded the Zeeman data set to include H~I, OH, and CN surveys \citep{2004ApJS..151..271H, 1999ApJ...520..706C, 2008ApJ...680..457T, 2008A&A...487..247F}. Their model for $B_{TOT}$ vs. $n_H$ had $B_{TOT,max} = B_0$ at lower densities, based on the most probable result from \cite{2005LNP...664..137H}. For higher densities the maximum $B_{TOT}$ had a power-law dependance, $B_{TOT,max} = B_0 (n/n_0)^\kappa$. The PDF of $B_{TOT}$ at each density was assumed to be flat, with the $B_{TOT}$ equally distributed between the $B_{TOT,max}$ at that $n_H$ and a lower limit $B_{TOT} = f \times B_0$, with $0 \le f \le 1$. A delta function PDF (all $B_{TOT}$ at each $n_H$ being the same) would have $f = 1$, while $f =0$ would be the flat PDF between $B_{TOT,max}$ and 0. The results for the four free parameters in the Bayesian model (figure~\ref{Bvsn}) were $B_0 \approx 10$ $\mu$G, $n_0 \approx 300$ cm$^{-3}$, $\kappa \approx 0.65$, and $f \approx 0$. 

For $n_H > n_o$ interstellar magnetic field strengths increase with density. Possible explanations are that diffuse clouds form by accumulation of matter along magnetic field lines, which would increase the density but not the field strength, or that there is a physical process such as turbulent magnetic reconnection that acts to keep fields from increasing with density \citep{1999ApJ...511..193V,2012ApJ...757..154L}. Once densities become large enough for clouds to be self-gravitating, gravitational contraction with flux freezing may lead to the increase in field strength with increasing density. 

The Bayesian analysis of the PDFs of the total field strength leads to the same result for the importance of magnetic fields with respect to gravity that was discussed above: for lower densities (where clouds are predominately not self-gravitating), the mass-to-flux ratio is subcritical. At higher densities it is supercritical.

The statistical increase in field strengths with density, parameterized by the power law exponent $\kappa$, may be compared with theoretical predictions. The ambipolar diffusion theory has $\kappa$ near zero at early stages when contraction of neutrals increases density but not field strengths; as evolution proceeds, $\kappa$ gradually increases to a maximum of $~0.5$, e.g., \cite{1999ASIC..540..305M}. The Bayesian analysis value of $\kappa \approx 0.65 \pm 0.05$ does not agree with the ambipolar diffusion prediction. It does agree with the value $\kappa = 2/3$ found by \cite{1966MNRAS.133..265M} for a spherical cloud with flux freezing. However, while spherical collapse does produce $\kappa = 2/3$, finding that clouds have $\kappa$ near this value does not require that clouds be spherical. It only means that collapse is approximately self-similar. The Bayesian result does imply that magnetic fields in self-gravitating clouds are generally too weak to dominate gravity in a large fraction of molecular clouds. However, the Bayesian analysis is a statistical one that does not rule out ambipolar diffusion being dominant in a small proportion of molecular clouds. 

\cite{2015MNRAS.451.4384T} have questioned the results of the Bayesian analysis described above on several grounds, including: i) that the clouds are not observed to be spherical; ii) that the Bayesian analysis included both H~I and molecular cloud data; a non-Bayesian analysis by \citet{2015MNRAS.451.4384T} of molecular cloud detections only yielded $\kappa \approx 0.5$; and, iii) that they found inferred cloud densities in a separate literature search often differing from those used by \citet{2010ApJ...725..466C}, particularly higher CN cloud densities, and argued that the CN points in figure~\ref{Bvsn} should move further right thus lowering $\kappa$. Collectively, these are open questions for which countervailing arguments and considerations exist; both are important to our full understanding of the scientific interpretation of Zeeman observations. On (i) it can be argued that real clouds invariably have significantly non-spherical morphologies due to other forces such as bulk flows and turbulence. Regarding (ii), omitting clouds with Zeeman non-detections (and accordingly smaller inferred magnetic field strengths) in a non-Bayesian analysis can bias the estimation of $\kappa$ downwards; the subset of clouds with larger field strengths may well have a smaller $\kappa$ than the total set. On the final point (iii), it is required to estimate the density of the Zeeman tracer as opposed to the highest density for each cloud. Further, high excitation lines of other molecular species may sample higher densities that the N=1-0 CN transition due to excitation and astrochemical depletion. As current and future telescopes provide further data, as described in Section \ref{sec_sum}, these questions will undoubtedly be further constrained.

\subsubsection{Radial Dependence of Mass/Flux}

Study of $M/\Phi$ such as that illustrated by figure~\ref{BvsN} compare different clouds. Also of interest is the variation of $M/\Phi$ within a cloud, for that can be indicative of the role of the magnetic field in the structure and evolution of a cloud. This is a very difficult observational task because spectral lines will generally be weaker away from cloud centers. However, \cite{2009ApJ...692..844C} reported such a study toward four dark clouds. Although determination of actual values of $M/\Phi$ requires knowledge of the unknown angle $\theta$ between the magnetic field vector and the line of sight, it is possible to map the variation from point to point within a cloud if one assumes that the magnetic field direction is the same at the various positions. This is a reasonable assumption if the magnetic field is strong and dominates turbulence, as in the standard ambipolar diffusion model of star formation. That model requires that $M/\Phi$ increase from envelope to core as collapse of neutrals through the magnetic field increases the mass but not (so much) the field strength in the core.

The \cite{2009ApJ...692..844C} result was that in all four clouds, $M/\Phi$ decreases from envelope to core -- the opposite of the ambipolar diffusion prediction. This observational result agreed with results from a weak field, turbulence dominated simulation \citep{2009ApJ...702L..37L}. The observed result could also be due to magnetic reconnection \citep{2005AIPC..784...42L}, since loss of magnetic flux due to turbulent reconnection will proceed more rapidly in envelopes that in cores, since in envelopes have larger spatial scales and in general stronger turbulence. 

\cite{2009MNRAS.400L..15M, 2010MNRAS.409..801M} reviewed the above results and conclusion, and argued that (1) motion of cores through surrounding more diffuse gas could lead to {\bf B} in cores and their envelopes not being essentially parallel and (2) that since $B_{LOS}$ was not detected in the envelopes only upper limits should be considered. \cite{2010MNRAS.402L..64C} discussed these arguments. The first point may have some validity, but observed correlation of $B_{POS}$ directions in cores and surrounding gas argues against it. In any case, such a process would sometimes increase and sometimes decrease the observed radial dependence of $M/\Phi$. Four clouds is not a large number, but all four did show the same result. On the second point, it is certainly true that at the $3\sigma$ upper-limit level, $M/\Phi$ constant or even decreasing slightly with radius is consistent with the data for each cloud individually, but the probability that this is true for all four clouds is $\sim3 \times 10^{-7}$. None the less, clear observational evidence for the ambipolar diffusion theory was not provided by the results in \cite{2009ApJ...692..844C}.

\subsubsection{Models of Specific Clouds}

Ambipolar diffusion models for specific clouds, B1 and L1544, have been produced for comparison with observational data including OH Zeeman detections \citep{1994ApJ...427..839C, 2000ApJ...529..925C}. In both cases the models could agree with observations, but both required that the fields be mainly in the plane of the sky, since the field strengths required by the models were much larger than the line-of-sight strengths obtained from Zeeman observations. While this could be true for the very small sample of two, in the larger sample of dark clouds with OH Zeeman observations one might expect to find examples of the field lying mainly along the line of sight, such that very large $B_{LOS}$ would be found from Zeeman observations. Such large fields are not found.

\section{Zeeman Observational Results - Masers}

Astrophysical maser components, due to their intrinsic nature as compact objects of high brightness temperature, are critical probes of magnetic fields in intermediate- and high-mass star forming regions (HMSFR) and are of unique importance in the study of the magnetic field over spatial scales of 10-100 AU \citep{surcis_evn_2013, vlemmings_magnetic_2010}. Hydroxyl (OH), water (H$_2$O), and methanol (CH$_3$OH) maser species have widespread association with HMSFR; each probes different physical conditions in these regions. SiO masers are rare toward SFR \citep{1992ASSL..170.....E}, and we do not discuss them in this review. Broad reviews of maser observations of star forming regions (SFR) are provided in the monographs by \citet{1992ASSL..170.....E} and \citet{2012msa..book.....G}. Polarization-specific observations of masers toward SFR are reviewed by \citet{vlemmings_maser_2008} and \citet{vlemmings_maser_2012}. In this article we seek to synthesize the current status of maser polarization observations of SFR, the impact of such observations on magnetic field estimates in these regions, and their relationship to open questions in star formation theory. All magnetic field values cited in this section are $B_{LOS}$ unless otherwise specified.

\subsection{OH masers}

Hydroxyl masers are common in SFR and are believed to lie in the enclosing dusty molecular envelope, arising during the development of the associated ultra-compact HII (UCHII) region and within the period when the UCHII is within approximately 30 milliparsec in size \citep{caswell_maser_2001}. Several sources are known to be somewhat larger in extent including OH 330.953-0.182 \citep{caswell_lba_2010} and OH 337.705-0.053 \citep{caswell_maser_2011}, the latter source perhaps approaching the end of its evolutionary maser-emitting phase.

\subsubsection{Main line OH masers}

OH maser emission toward SFR is detected most frequently in the ground-state main line transitions ${}^2\Pi_{\frac{3}{2}},J=\frac{3}{2},\{F=1\rightarrow 1, F=2\rightarrow 2\}$ at 1665 MHz and 1667 MHz respectively and in the excited-state transitions ${}^2\Pi_{\frac{3}{2}},J=\frac{5}{2},\{F=2\rightarrow 2, F=3\rightarrow 3\}$ at 6031 MHz and 6035 MHz respectively. The Zeeman effect is readily detected in OH maser transitions due to the paramagnetic nature of the hydroxyl radical \citep{1977cema.book.....C,1992ASSL..170.....E}. In the formalism of the foundational theory of maser polarization \citep{goldreich_astrophysical_1973} the Zeeman splitting will exceed the maser linewidth if the magnetic field $B > 0.5$mG \citep{slysh_total_2002}. The Zeeman pattern is as discussed above; as noted in that discussion, fully-separated Zeeman components allow inference of the total magnetic field $B_{TOT}$. 
The $\pi$ components are not frequently observed \citep{slysh_total_2002} but are not completely absent; \citet{green_excited-state_2015} find an incidence of $\sim 16\%$ in excited state OH transitions.

OH Zeeman pairs are frequently detected toward SFR in interferometric observations sensitive to circular or full polarization. In this paragraph we consider such observations of the main line 1665 MHz and 1667 MHz OH masers. The contemporary MAGMO survey of the Carina-Saggitarius tangent in these OH transitions toward methanol maser sites and previously-known OH sources using the Australia Telescope Compact Array (ATCA) by \citet{green_magmo:_2012} detected 11 Zeeman components and found OH maser fractional linear polarization $m_l \sim 22-95\%$ and fractional circular polarization $m_c \sim 6-100\%$. These observations and prior aggregated OH Zeeman measurements in this region span a B-field range: -1.5mG $<$ B $<$ +8.9 mG. Interferometric studies of individual sources find broadly comparable magnetic field magnitudes, including Very Long Baseline Array (VLBA) observations of G23.01-0.41 \citep{sanna_vlbi_2010} and W75(N) \citep{slysh_total_2002}, Long Baseline Array (LBA) observations of OH 337.705-0.053 \citep{caswell_maser_2011}, OH 330.953-0.182 \citep{caswell_lba_2010}, OH 300.969+1.147 \citep{caswell_maser_2009}, and 323.459-0.079 \citep{caswell_maser_2001-1}, and MERLIN observations of IRAS 20126+4104 \citep{edris_masers_2005}. A recent extensive Parkes single-dish polarization spectroscopic survey found that approximately one third of the main-line OH masers toward SFR have a feature that is at least 50\% linearly polarized \citep{caswell_parkes_2013}. Single-dish observations with the Nan\c{c}ay Radio Telescope (NRT) detected several Zeeman features with inferred magnetic fields consistent with the interferometric results cited above \citep{2014ARep...58..462B}.

\subsubsection{Excited state OH masers}

Excited-state OH masers at 6 GHz are usually strongly associated with 1665 MHz OH masers toward SFR, although at perhaps one third the incidence to the same sensitivity level \citep{caswell_maser_2001}. \citet{caswell_oh_2004} note that this is consistent with pumping models predicting similar conditions \citep{2000ApJ...534..770P,2002MNRAS.331..521C}. Ground-state OH masers have representative dust temperatures $\ge$ 100 K, gas temperatures $\le$ 100 K, and density  $10^4 < n_H < 10^{8.3}$ cm$^{-3}$ \citep{2002MNRAS.331..521C}. Excited-state OH masers trace somewhat cooler gas, at higher density $10^{6.5} < n_H < 10^{8.3}$ cm$^{-3}$ \citep{2002MNRAS.331..521C,green_merlin_2007}.

In this paragraph we consider recent interferometric Zeeman or full polarization observations of 6 GHz excited state OH maser emission toward SFR. A contemporary survey of 30 source positions using the ATCA by \citet{green_excited-state_2015} detected 94 Zeeman pairs and 18 Zeeman triplets with inferred magnetic fields -10.4 mG $< B <$ 11.4 mG. Interferometric observations of individual sources find Zeeman pairs with a comparable range of magnetic field magnitudes including European VLBI Network (EVN) observations of W3(OH) \citep{fish_structure_2007}, MERLIN observations of W51 \citep{etoka_methanol_2012}, W3(OH) \citep{etoka_association_2005}, and ON1 \citep{green_merlin_2007}, ATCA observations of OH 353.410-0.360 \citep{caswell_maser_2001-1}, LBA observations of G351.417+0.645 and G353.410-0.360 \citep{caswell_magnetic_2011}, and Very Large Array (VLA) observations of NGC 6334I \citep{hunter_extraordinary_2018}.

\subsubsection{1720 MHz OH masers}

The ${}^2\Pi_{\frac{3}{2}},J=\frac{3}{2}, F=2\rightarrow 1$ OH maser transition at 1720 MHz is less frequently observed and detected toward SFR. They are believed to be approximately 1/6th as prevalent as 1665 MHz OH masers toward SFR when surveyed to the same sensitivity limit \citep{caswell_oh_2004}. Observationally 6035 MHz and 1720 MHz OH masers are known to have correlated association \citep{caswell_maser_2001}; at high resolution \citet{fish_structure_2007} find components in these transition to be co-spatial within 10-20 mas. An ATCA interferometric survey of 1720 MHz OH masers associated with 1665 MHz and 6035 MHz OH maser sites found Zeeman pairs with associated magnetic field magnitudes as high as $\sim 16$ mG \citep{caswell_oh_2004}. These authors argue that the 1720 MHz OH masers toward SFR trace regions of higher densities and higher associated magnetic fields $(\sim 1.5-2\times$) accordingly.

\subsubsection{Detection of Zeeman pairs}

OH Zeeman pairs can be difficult to identify unambiguously due to flux density or positional offsets between the two $\sigma$ components \citep{1977cema.book.....C}, likely due to differing maser amplification paths for the two separated components. This effect is more pronounced for the main line OH maser transitions as they have larger Zeeman splitting coefficients $Z$ than the 1720 MHz OH transition and 6 GHz excited state OH transitions. These values are 0.113 km/s/mG at 1720 MHz \citep{caswell_oh_2004}, 0.079 km/s/mG at 6030 MHz and 0.056 km/s/mG at 6035 MHz \citep{caswell_magnetic_2011}; cf. table \ref{Zs}.

\subsection{Methanol masers}

Methanol masers have emerged as particularly powerful probes of star formation. The 6.7 GHz $5_1 \rightarrow 6_0 A^{+}$ methanol transition is associated only with HMSFR \citep{2003A&A...403.1095M, green_merlin_2007}. It is difficult to infer local magnetic field structure in high-resolution observations of ground-state OH masers toward SFR due to significant external Faraday rotation \citep{surcis_methanol_2009}; methanol maser transition frequencies are far less affected.

Methanol masers are classified as either Class I or Class II \citep{1991ASPC...16..119M,1991ApJ...380L..75M}. Class I methanol masers trace shocked gas at the interfaces of outflows from HMSFR \citep{2009ApJ...702.1615C}. A comprehensive review of Class I methanol masers and their excitation is provided by \citet{leurini_physical_2016}. Class II methanol masers are found closer to massive protostars within HMSFR. \citet{wiesemeyer_polarization_2004} cites W3(OH) as the prototype Class II methanol maser source. The two strongest Class II methanol maser transitions are the $2_0 \rightarrow 3_{-1} E$ 12.2 GHz transition \citep{1987Natur.326...49B} and the $5_1 \rightarrow 6_0 A^{+}$ 6.7 GHz transition \citep{1991ApJ...380L..75M}. Typical physical conditions for Class II methanol maser excitation are cited by \citet{wiesemeyer_polarization_2004} as $n \sim 10^{5-8}$ cm$^{-3}$ with a gas temperature less than the dust temperature (both $\le 100$ K).

The methanol molecule is non-paramagnetic and expected to have low linear and circular polarization in an external magnetic field \citep{green_merlin_2007,vlemmings_new_2008} particularly if partially saturated, as described in the general maser theory summarized by \citet{watson_magnetic_2009}. A further complication for Zeeman polarimetry of methanol masers was immediately presented once these observations became technically feasible: no accurate laboratory measurement existed for the Land\'{e} g-factor for the transitions of interest. The community relied on an uncertain extrapolation of laboratory measurements of 25 GHz methanol transitions \citep{1951PhRv...81..197J}, and the extrapolation calculation may have been in error by an order of magnitude \citep{vlemmings_zeeman_2011}. Recently a full quantum mechanical derivation has been performed \citep{lankhaar_characterization_2018} resulting in an accurate Zeeman coefficient. Accordingly we do not cite inferred magnetic fields from work using earlier Land\'{e} g-factors than \citet{lankhaar_characterization_2018}, but only note Zeeman velocity or frequency splitting for those results accordingly.

\subsubsection{6.7 GHz methanol masers}

In this paragraph we confine our discussion to polarization observations of the 6.7 GHz methanol maser transition toward HMSFR. \citet{2002IAUS..206..151E} reported the detection of linear polarization at levels of up to 10\% toward NGC6334F using the ATCA. MERLIN observations of W3(OH) by \citet{vlemmings_methanol_2006} detected a median linear polarization fraction $m_l \sim 1.8\%$ and set an upper limit to the fractional circular polarization $m_c < 2\%$; the upper limit to the Zeeman splitting was $v_z < 1.1\times 10^{-3}$ km/s. \citet{dodson_first_2008} reported linear polarization $m_l \sim 0.5-10.1\%$ toward G339.88-1.26 using the LBA. The first systematic survey for Zeeman components was performed by \citet{vlemmings_new_2008} using the Effelsberg single-dish telescope to observe a sample of 24 sources; 17 Zeeman components were detected with an average Zeeman splitting of 0.56 m/s. Figure \ref{Masers} shows an example of a Zeeman detection Stokes I and V profile. Zeeman splitting toward the period flaring source G09.62+0.20 was reported by \citet{vlemmings_possible_2009} using similar Effelsberg observations. As more telescopes and interferometer arrays became equipped with receivers in this band the scope of observations of this transition toward HMSFR increased significantly. MERLIN observation of DR21(OH) and DR21(OH)N were completed by \citet{harvey-smith_first_2008}, of Cepheus A HW2 by \citet{vlemmings_magnetic_2010}, and of IRAS 18089-1732 by \citet{dallolio_methanol_2017}. \citet{dodson_probing_2012} undertook a survey of ten SFR using the ATCA; and the Mount Pleasant 26-m single dish telescope was equipped for polarimetry in this transition by \citet{stack_polarisation_2011}, who detected Zeeman splitting in the periodic flaring source G9.62+0.20. A significant systematic survey campaign of HMSFR using the EVN is reported by \citet{surcis_methanol_2009,surcis_evn_2012,surcis_evn_2013,surcis_evn_2015}; these authors report Zeeman $v_z$ splitting spanning the range from -9.7 m/s to +7.8 m/s and fractional linear polarization in the range $m_l \sim 0.4-17\%$. We do not imply that the data are uniformly distributed in these ranges; an approximate summation here suggests a mean unsigned $|v_z|\sim3.2$ m/s and a mean fractional linear polarization ${\bar m}_l \sim 3.6\%$. \citet{vlemmings_zeeman_2011} cite a fitted Zeeman splitting dispersion $\left<\triangle V_Z\right> \sim 0.62$ m/s from their Effelsberg single-dish Zeeman survey. \citet{lankhaar_characterization_2018} report that this translates to a mean magnetic field magnitude $\sim 12$ mG using the correct Zeeman splitting coefficient.

\subsubsection{36 GHz and 44 GHz methanol masers}

The Class I $4_{-1}\rightarrow 3_0 E$ 36 GHz methanol maser transition was detected toward the HMSFR M8E using the VLA by \citet{sarma_detection_2009}; these authors resolved two Zeeman components with splitting $v_z=34.4 \pm 5.9$ Hz and $v_z=-53.2 \pm 6.0$ Hz respectively. Zeeman splitting has also been detected in the Class I $7_0 \rightarrow 6_1 A+$ 44 GHz methanol maser transition toward the SFR OMC-2 \citep{sarma_discovery_2011,momjian_comparison_2012} and DR21(OH) \citep{momjian_zeeman_2017}. In DR21(OH) the authors find $v_z=53.5\pm 2.7$ Hz and for OMC-2 report $v_z=18.4\pm 1.1$ Hz and $v_z=17.7 \pm 0.9$ Hz over two epochs. 

Using the correct Land\'{e} g-factors computed for these transitions and the most likely hyperfine transitions, \citet{lankhaar_characterization_2018} infer a magnetic field magnitude of 20-75 mG for the reported 36 GHz and 44 GHz methanol maser Zeeman observations. \citet{momjian_zeeman_2017} note that these masers sample densities $n \sim 10^{7-8}$ cm$^{-3}$ \citep{leurini_physical_2016}. As noted by \citet{lankhaar_characterization_2018} the B-field magnitudes $\sim20-75$ mG are not inconsistent with shock compression.

\subsubsection{Other millimeter wavelength transitions}

\citet{wiesemeyer_polarization_2004} conducted single-dish polarimetry of methanol masers at higher frequencies using the IRAM 30-m telescope. They observed the Class I transitions: $5_{-1} \rightarrow 4_0 E$ (85 GHz), $8_0 \rightarrow 7_1 A+$ (95 GHz), $6_{-1} \rightarrow 5_0 E$ (132 GHz); and Class II transitions: $3_{1}\rightarrow 4_0A+$ (107 GHz) and $6_0 \rightarrow 6_{-1} E$ (157 GHz). The authors report fractional linear polarization in the range $m_l \sim 2.8-39.5\%$ and fractional circular polarization $m_c \sim -7.1 - 3.52\%$. They argue that the masers are unsaturated and that the theory of \citet{1992ApJ...384..185N} applies; specifically that the linear polarization arises from anisotropic pumping and anisotropic radiation losses and the circular polarization from non-Zeeman effects \citep{watson_magnetic_2009}. \citet{kang_linear_2016} present a survey of linear polarization in 44 GHz and 95 GHz Class I masers conducted with a single dish from the Korean VLBI Network (KVN). Their results yielded a lower fractional linear polarization and they argue accordingly for a diminished influence of anisotropic pumping and losses than \citet{wiesemeyer_polarization_2004}.

\subsection{Water masers}

The advent of the VLBA was accompanied by the development of observing techniques that allowed VLBI polarimetry of 22 GHz water masers \citep{1998ApJ...507..909L}. VLBI observations of the Zeeman effect in 22 GHz water masers were first reported by \citet{2006A&A...448..597V} and \citet{2008ApJ...674..295S}; such observations allow the magnetic field in SFR to be measured in high-density regions $n \sim 10^9$ cm$^{-3}$. Contemporary VLBI observations of W75N are reported by \citet{surcis_structure_2011}. Their analysis leads to an inferred magnetic field of ~200 mG to 1000 mG in the shocked outflow region where the water masers arise. Modeling suggests that the shock is a C-shock. As part of their comprehensive observations of the HMSFR G23.01-0.41 \citet{sanna_vlbi_2010} included VLBA observations of H$_2$O masers, confirming that the water masers trace fast outflows from the massive YSO in this source.

\subsection{Relationship of maser polarization observations to star formation theory}

There remain broad open questions in both the micophysics and macrophysics of star formation theory \citep{2007ARA&A..45..565M}. Maser polarization observations in SFR probe the fine-scale magnetic field in dense regions $n\sim10^{5-11}$ cm$^{-3}$ of obscured HMSFR \citep{kang_linear_2016} and therefore provide constraints on the theory of high-mass as opposed to low-mass star formation  \citep{zinnecker_yorke_2007}. In addition, masers are confined to certain evolutionary phases of HMSF, specifically between the formation of hot dense molecular cores (HDMC) \citep{zinnecker_yorke_2007}, equivalently high-mass protostellar objects (HMPO) \citep{2007ARA&A..45..565M}, and their subsequent evolution into ultra-compact HII regions (UCHII) \citep{churchwell_2002}. The hot molecular core phase is associated with the emergence of outflows, jets, and water and methanol maser emission \citep{2007ARA&A..45..565M,zinnecker_yorke_2007} with the later emergence of OH maser emission during the development of an UCHII region \citep{churchwell_2002}. HMPOs have high extinction, complex spatial structure due to clustered HMSF, and may lie at large distances \citep{2007ARA&A..45..565M}. Maser observations can mitigate these observational challenges however and provide unique probes of the fine-scale kinematics and magnetic fields in HMPO regions. However it is critical to locate the maser emission within the broader kinematic and dynamical picture of high-mass star formation regions as accurately as possible in order to assess the physical meaning of the magnetic field measurements inferred from maser observations and their overall role as probes of HMSF in general.

Class II methanol masers lie closer to the central HMPOs but there has been debate as to their exact location and kinematics; we consider the available evidence here. Early interferometric imaging of 6.7 GHz and 12.2 GHz methanol masers toward SFR revealed frequent linear or arc-like features, which were interpreted as Keplerian disks \citep{1993ApJ...412..222N,1998ApJ...508..275N}. However, as described by \citet{dodson_probing_2012} competing interpretations have arisen, including shocks \citep{1998MNRAS.301..640W}, disk infall or outflow \citep{2009A&A...502..155B}, and shock interaction with rotating molecular clouds \citep{2004MNRAS.351..779D}. \citet{dodson_probing_2012} conducted a survey of 10 SFR in the 6.7 GHz methanol transition using the ATCA in order to distinguish these cases by comparing the gross polarization morphology and the structural morphology of the maser distribution; this statistical test was inconclusive. Using a larger sample of SFR observed in full polarization in the 6.7 GHz methanol transition using the EVN, \citet{surcis_evn_2013, surcis_evn_2015} have synthesized all external information on outflow direction from other molecular or dust polarization observations for their sample and searched for correlations between structural maser distribution and polarization EVPA and outflow direction. They find a statistically significant correlation supporting the alignment of the inferred magnetic field direction and the large-scale outflow direction. This is not universal but is clearly indicated in several sources studied in detail in the 6.7 GHz methanol transition. Cepheus A HW2 shows methanol masers in an elliptical ring of size $\sim 650$ AU that probe material being accreted onto the disk \citep{vlemmings_magnetic_2010}. There is no clear sign of rotation and the magnetic field is aligned with the outflow. The HMSFR G23.01-0.41 observed by \citet{sanna_vlbi_2010} using VLBI shows methanol masers in a toroid consistent with expansion and rotation about a massive YSO; this is confirmed in an associated proper motion analysis. MERLIN observation of W51 by \citet{etoka_methanol_2012} show the methanol masers around e2-W51 to be in a velocity-coherent structure perpendicular to the CO outflow; the maser spots appear to trace the magnetic field lines elsewhere in the source. \citet{harvey-smith_first_2008} find a methanol maser distribution toward DR21(OH) consistent with a Keplerian disk toward DR21(OH)N. The preceding results show that methanol masers are unique probes of disks or toroids surrounding HMPOs and that there is significant potential in future observations. These inner regions are critical to understanding key issues in HMSF including accretion mechanisms \citep{krumholz_2007}, angular momentum transport, disks \citep{cesaroni_2007} and outflows. The physics of low- and high-mass star formation differ in key respects, as summarized by \citet{zinnecker_yorke_2007}.

In several sources it is particularly clear that the measured methanol maser magnetic fields sample an overall coherent magnetic field and not isolated regions of dense shocked material. This is argued for Cepheus A HW2 by \citet{vlemmings_magnetic_2010} who also cite additional supporting evidence for the source W75N \citep{surcis_methanol_2009}. There is a particularly strong alignment between the methanol maser magnetic field orientation and the field derived from dust polarization in W51-e2 \citep{surcis_evn_2012}; in particular see their Figure 10. The evidence that maser polarization observations sample the global field is particularly valuable constraining HMSF collapse, accretion, and angular momentum transport mechanisms \citep{2007ARA&A..45..565M}. 

\subsection{Magnetic fields and density relation}

Using Zeeman magnetic field measurements made possible for different maser species in the past decade, particularly with the addition of methanol and water masers, \citet{vlemmings_new_2008} compared the resulting magnetic field - density relation for Cepheus A against the relation $B \propto n^{0.47}$ derived by \citet{1999ApJ...520..706C} finding excellent agreement; see figure \ref{Masers}. This is confirmed after revisions accounting for the more accurate Land\'{e} g-factor for the methanol molecule \citep{lankhaar_characterization_2018}. The agreement with this relation has been used as an independent consistency check on measurements of maser magnetic fields \citep{momjian_zeeman_2017,surcis_structure_2011}. However, note that the \citet{1999ApJ...520..706C} study has been superceded by a more complete and statistically improved study that yields an exponent of $\approx 0.65$. With the uncertainties in especially densities sampled by masers, in addition of other uncertainties discussed above, the \citet{vlemmings_new_2008} result is not inconsistent with the higher exponent. But it is also true that the physics governing the relationship between field strength and density may be completely different in the extended gas and the regions with the special conditions that give rise to masers; there is no strong astrophysical argument that the same exponent hold in both regions. 

Future observations offer the potential of extending measurement of the magnetic field - density relation at maser densities over a larger sample of individual sources. It is challenging to aggregate these data but new instrumental capabilities increasingly make this a possibility. A larger ensemble of sources will offset the inherent uncertainty in maser excitation density and provide tighter constraints on the $B \propto n^{\kappa}$ exponent in maser regions and better understanding of the underlying physics. We refer the reader to the discussion in Section 4.2 concerned the physical meaning of different exponent values.

\section{Summary and Conclusions}\label{sec_sum}

Interstellar Zeeman detections have been made in five species, H~I, OH, CN, 
CH$_3$OH, and H$_2$O, that sample densities n$_H$ over about 10 orders of magnitude ($10^{0-10}$ cm$^{-3}$). Hence, information about magnetic field strengths is available over the full range of densities from diffuse atomic clouds to very dense gas in regions of star formation. There are, however, a number of limitations in the data. Zeeman detections generally involve high sensitivity observations that require long telescope observation time, which limits the quantity of data available. In most cases the Zeeman effect provides only the field strength along the line of sight (with the exception of OH masers with fully separated Zeeman components), so statistical analysis is necessary. Obtaining astrophysical information such as the mass/flux ratio and the scaling of field strength with density requires knowledge of the column densities and volume densities traced by the Zeeman species. Particularly for masers, where the range of physical conditions under which masing may occur can be broad, this introduces significant uncertainties. Nonetheless, since the Zeeman effect provides the only direct technique for measurement of interstellar magnetic field strengths, Zeeman observations are crucial for our understanding of the role of magnetic fields in the evolution of interstellar clouds and in star formation.

At the lower densities sampled by H~I, field strengths do not appear to be systematically dependent on volume density, with maximum strengths of 10-20 $\mu$G. At densities starting at about 300 cm$^{-3}$, field strengths increase with density, following a scaling $B \propto n^{0.65}$, based on Bayesian analysis of OH and CN Zeeman data in extended gas. At higher densities, $\gtrsim 10^6$ cm$^{-3}$, results come from masers, and density estimates are less certain than for extended gas. Nonetheless, it is clear that field strengths continue to increase as a power law of density; early results suggested an exponent of 0.5 in this regime but on inspection current data are not inconsistent with a 0.65 exponent. Further analysis over a larger source sample is needed.

The observed ratio of column density to field strength is directly proportional to the mass/flux ratio $M/\Phi$, which measures whether gravitational or magnetic energy dominates. At lower densities $M/\Phi$ is subcritical; magnetic fields dominate gravity. Again at about densities of 300 cm$^{-3}$  the situation changes, and $M/\Phi$ becomes supercritical. 

These two measures of the relative importance of gravity and magnetic fields give the same result. At about the density at which clouds become self-gravitating, the importance of magnetic fields changes. The Zeeman data are consistent with a picture in which a cloud forms by flows along magnetic flux tubes (sometimes described as converging flows), increasing the local mass and density but not the field strength, until the cloud becomes magnetically supercritical and contracts gravitationally and eventually forms stars. Turbulent magnetic diffusion probably plays a large role in moderating field strengths during the evolutionary process. While this broad picture appears generally to satisfy data constraints, cloud evolution and star formation is a very complex process and this broad picture may not (always) be correct.

Although there have been extensive Zeeman observations in molecular clouds,  conclusions remain tentative. Additional observational work that should lead to more definitive conclusions are: (1) measurement of the overall $M/\Phi$ of molecular cloud complexes to study the degree of magnetic support; (2) additional Zeeman measurements at high densities in order to solidify $\kappa$ in the $B \propto \rho^\kappa$ relation; (3) additional studies of the magnetic field morphology and strength both within cores and between cores and between GMCs. 

Progress in interstellar Zeeman observations, particularly in non-masing lines, has slowed considerable in the last decade, due to the requirement of very large amounts of telescope time and instrumental polarization problems with important telescopes such as the IRAM 30-m and the ALMA telescopes that have prevented Zeeman observations in recent years. Hopefully these instrumental problems will be overcome so Zeeman data can be significantly extended in coming years. Being able to measure the Zeeman effect in protostellar disks with (for example) CN transitions would add significantly to information about magnetic fields. There has however been significant progress in maser observations of SFR in the past decade, particularly VLBI polarimetry sensitive to the Zeeman effect in receiver bands allowing observations of important tracers of HMSFR such as the 6.7 GHz methanol maser transition. This work has been facilitated both by improvements in observing and analysis techniques and advancement in the application of the theory polarized maser radiation transport. It has become clear that methanol masers in particular trace coherent magnetic field structures in HMSFR, provide important Zeeman probes of the magnetic field near massive YSOs, and add particular value to understanding star formation when considered with all other molecular and dust tracers. 

As noted above, the commissioning of circular polarization capability on ALMA has been technically challenging, however recent results are highly encouraging \citep{vlemmings_2017,vlemmings_2019}. The latter work presents an upper limit on the magnetic field in the disk of TW Hya based on, at present, the non-detection of the Zeeman effect in CN. The feasbility of Zeeman observations of CN emission in protostellar disks was considered earlier by \citet{brauer_2017}. A review of the potential of ALMA for linear polarization maser observations is provided by \citet{perez-sanchez_2013}; the linear polarization capabilities of ALMA have been realized much earlier in telescope operation due to their lower technical complexity than circular polarization for ALMA.
 
The ngVLA\footnote{https://ngvla.nrao.edu} science case for the study of magnetic fields in SFR is presented by \citet{hull_2018}. The science case includes Zeeman observations of a range of Galactic thermal and maser emission, including maser emission from the near environments of YSOs, OH masers to study the large-scale magnetic field distribution in the Milky Way, and Zeeman observations to detect magnetic fields in extragalactic OH masers sources. The ngVLA offers significantly increased sensitivity, a key consideration along with instrumental polarization purity in the technical feasibility of Zeeman observations, and nearly continuous coverage of the useful frequency spectrum from ~1 GHz to ~116 GHz. The potential for Zeeman observations using the Square Kilometre Array (SKA\footnote{https://www.skatelescope.org}) is described by \citet{robishaw_2015}. The high sensitivity of the SKA similarly enhances the technical feasibility of Zeeman observations; the authors note specific opportunities for significantly improved Zeeman observations to measure magnetic field densities in SFR with both OH and methanol masers, HI absorption studies toward background continuum sources as well as diffuse HI emission, OH masers tracing the Galactic magnetic field structure, and extragalactic masers. The SKA precursor projects MeerKAT\footnote{https://www.ska.ac.za} and ASKAP{http://www.atnf.csiro.au/projects/askap/index.html} have lower sensitivity
and reduced frequency coverage relative to the full SKA, but will make important contributions to the areas of SKA Zeeman science mentioned above.

\section*{Author Contributions}

RMC is primarily responsible for sections 1-4 and AJK for section 5. Section 6 is the responsibility of both.

\section*{Funding}

RMC receives support from NSF AST 18-15987. 
\section*{Acknowledgments}
RMC and AJK acknowledge with thanks the many colleagues with whom they have worked on Zeeman observations.

\bibliographystyle{frontiersinSCNS_ENG_HUMS} 
\bibliography{Zeeman}

\newpage

\begin{figure}[t!]
\includegraphics[bb= 40 180 380 540,clip,angle=-90,scale=1.0]{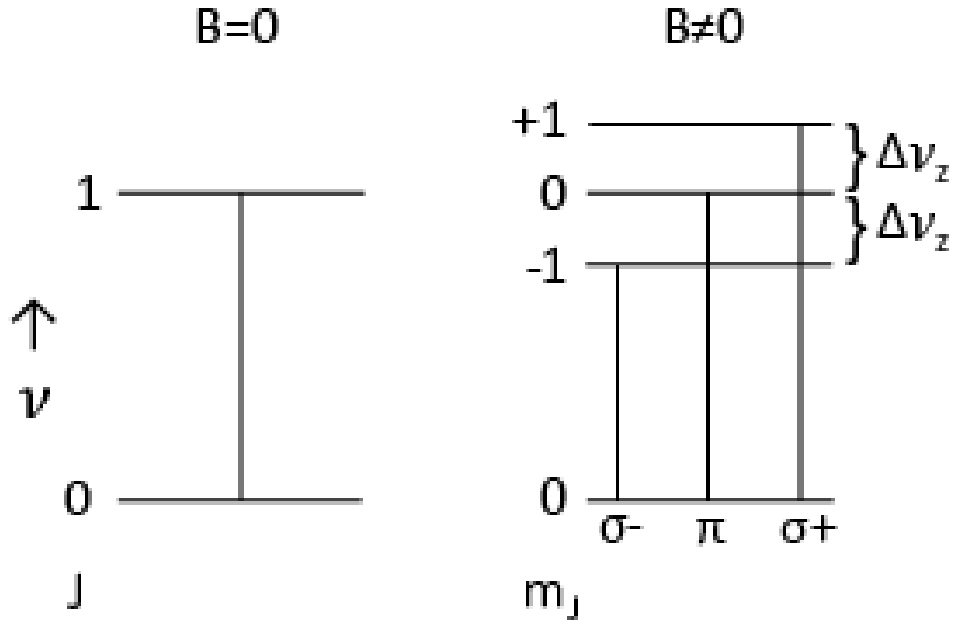}
\end{figure}
\begin{figure}[t!]
\includegraphics[origin=lb,bb= 50 0 730 380,angle=0,scale=0.7]{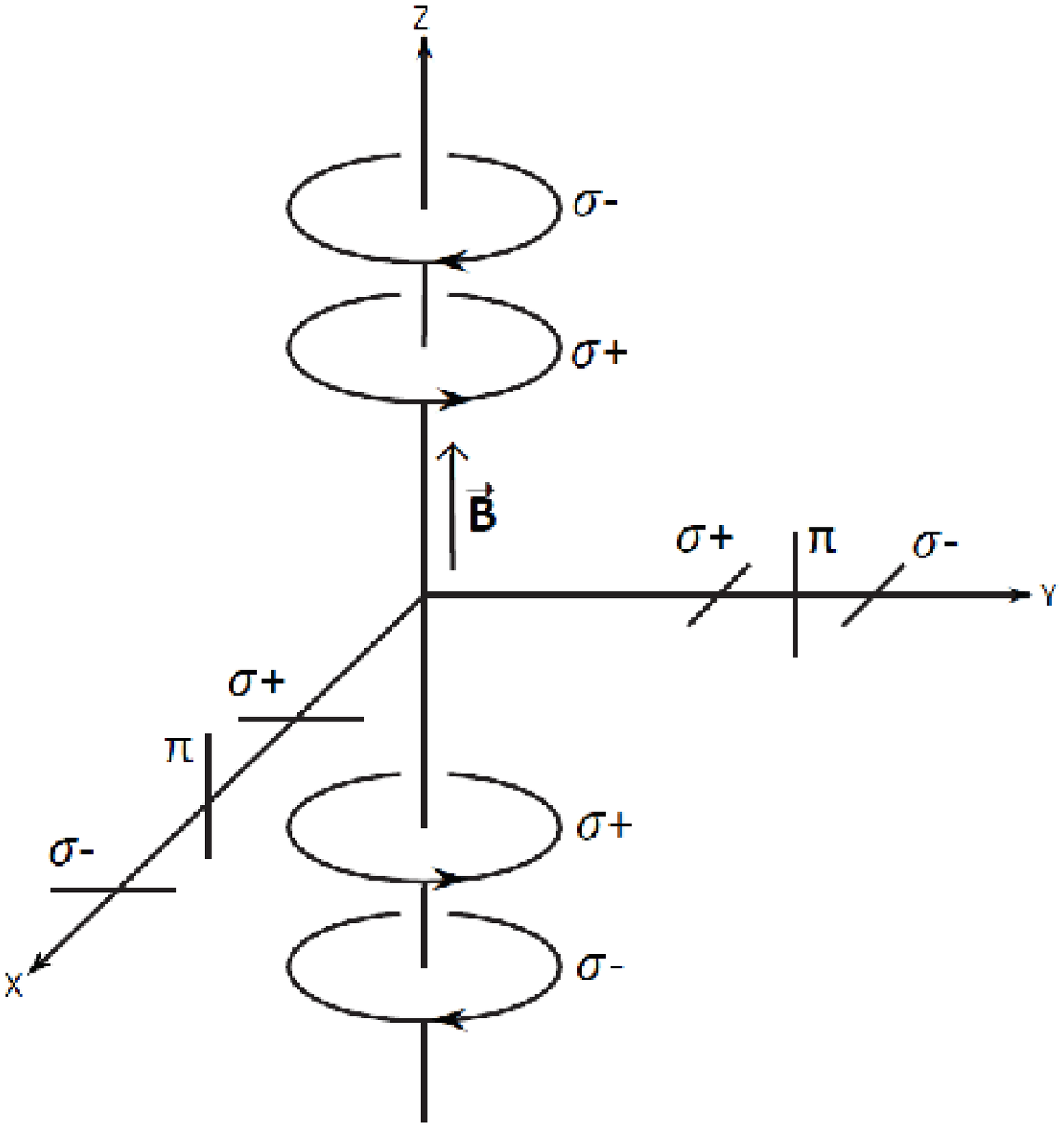}
\caption{Top: Energy level diagram showing Zeeman splitting. Bottom: Polarizations that would be observed from different angles with respect to the magnetic field.}
\label{ZeemanEnergy&Pol}
\end{figure}

\begin{figure}[t!]
\includegraphics[bb= 0 50 700 750,clip,angle=0,scale=0.9]{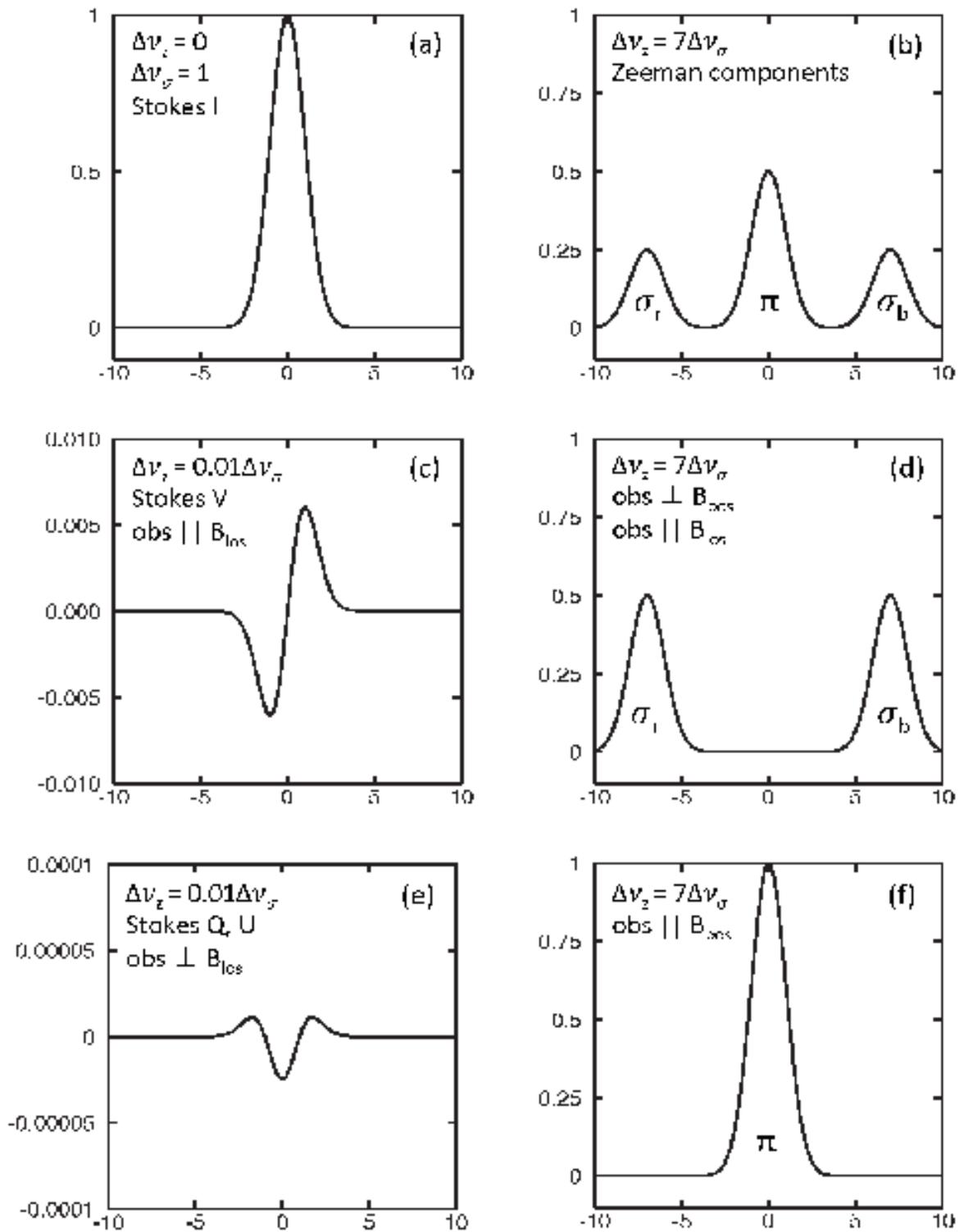}
\caption[Simulated Zeeman profiles]{Simulated Zeeman profiles for a Gaussian line profile. Units are relative with respect to peak of the Stokes I profile = 1.}
\label{ZeemanProfiles}
\end{figure}

\begin{figure}[t!]
\includegraphics[angle=0,scale=0.33]{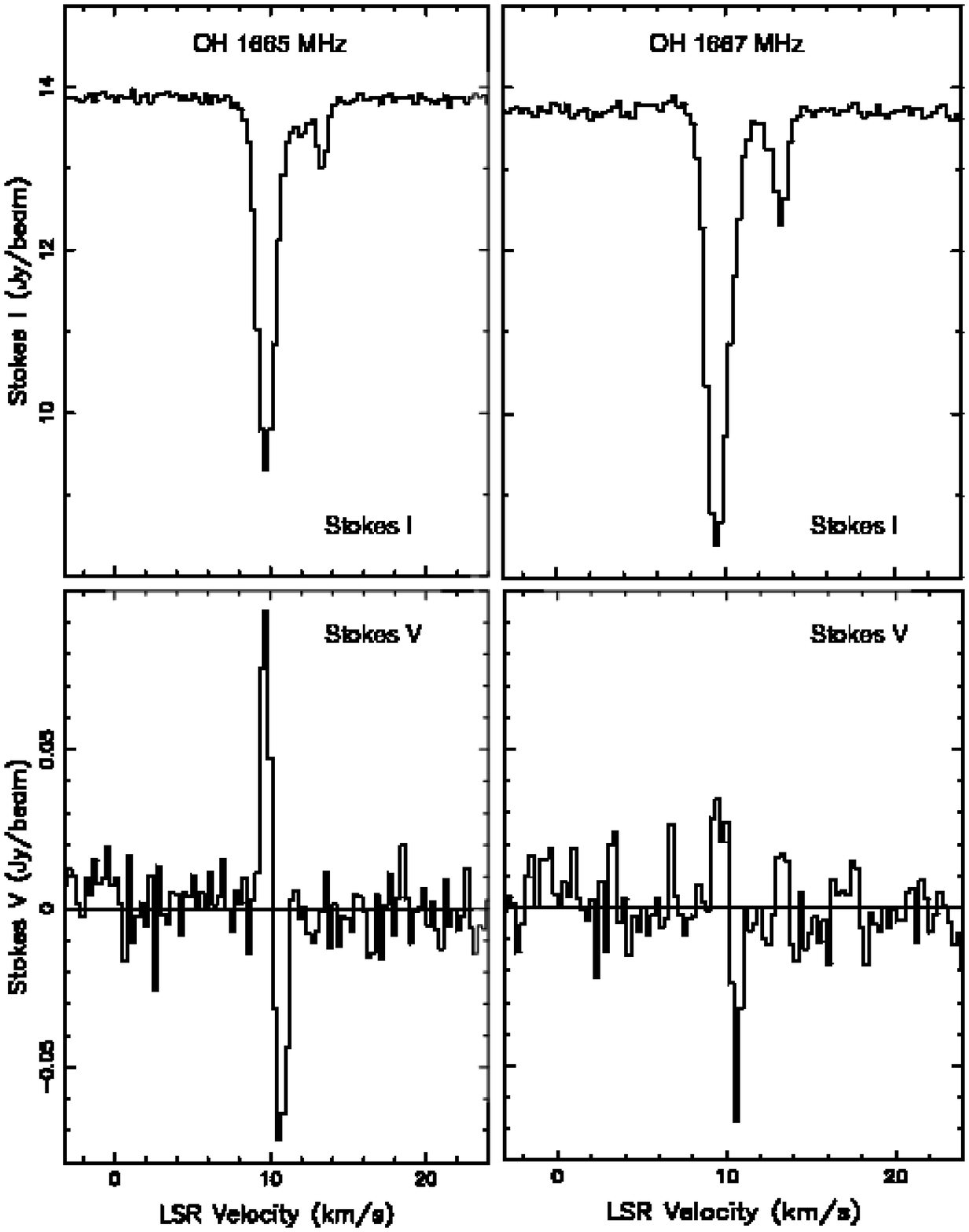} 
\includegraphics[origin=rb,bb= 0 0 490 620,clip,angle=-90,scale=0.48]{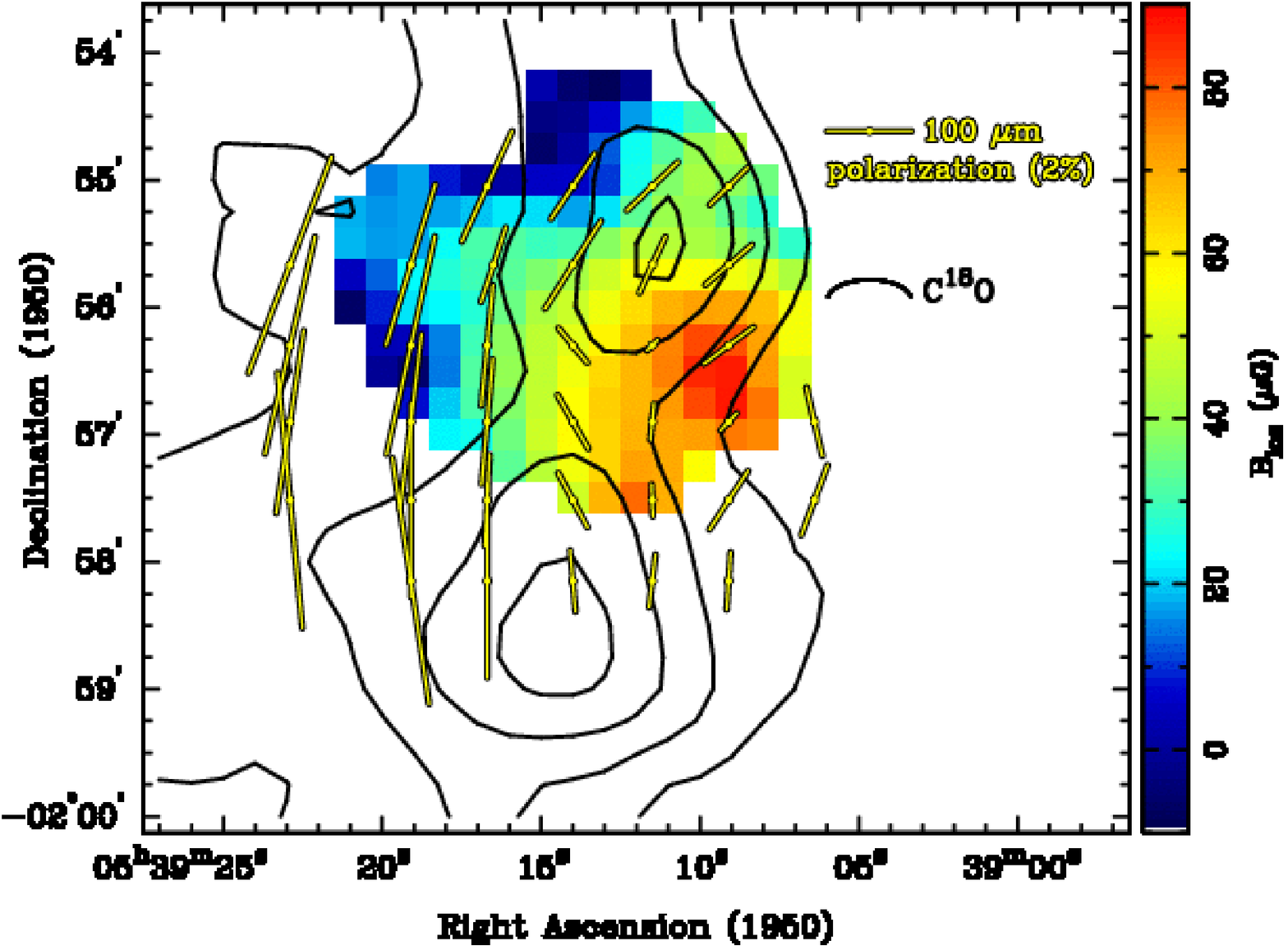} 
\caption[OH Zeeman VLA results for NGC2024]{Left: OH Zeeman Stokes I and V profiles toward NGC2024 at the peak $B_{LOS}$ position from VLA mapping \citep{1999ApJ...515..275C}. Right: Map of $B_{LOS}$ (color) from OH Zeeman. Contours are C$^{18}$O intensities and yellow line segments are dust polarization directions \citep{1995ASPC...73...97H}. The magnetic field in the plane of the sky is perpendicular to the dust polarization, hence roughly along the minor axis of the molecular cloud defined by C$^{18}$O (horizontal in the figure).}
\label{NGC2024}
\end{figure}

\begin{figure}[t!]
\includegraphics[scale=0.7]{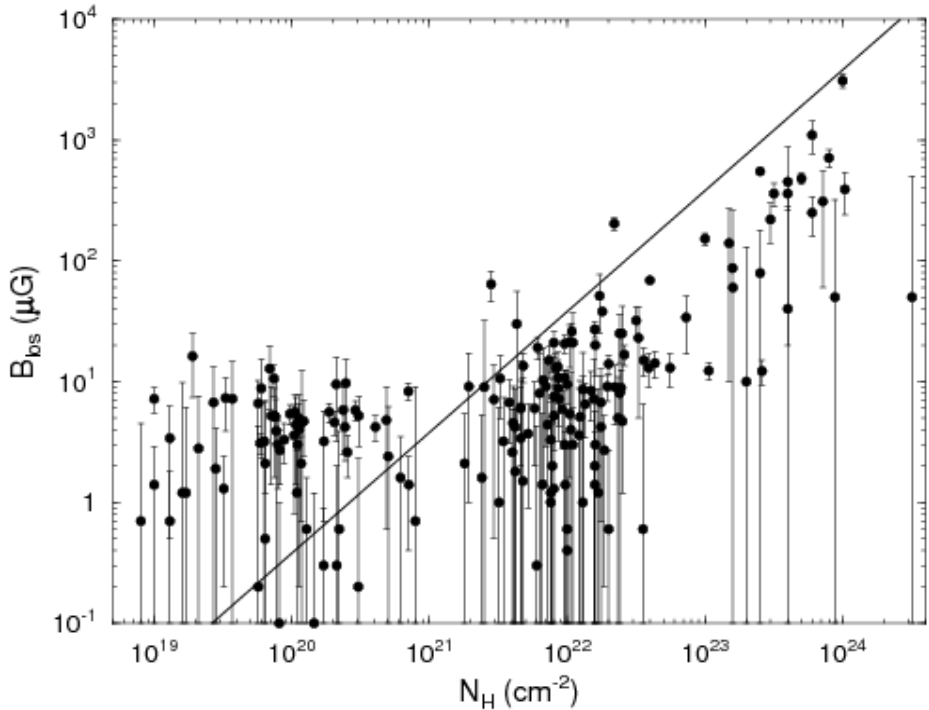}
\caption[Line-of-sight magnetic field strength versus column density]{H~I, OH, and CN Zeeman measurements of $B_{LOS}$ versus $N_H = N_{H~I} + 2N_{H_2}$. The straight line is for a critical $M/\Phi = 3.8 \times 10^{-21} N_H/B$. Measurements above this line are subcritical, those below are supercritical.}
\label{BvsN}
\end{figure}

\begin{figure}[t!]
\includegraphics[bb= 0 250 700 600,clip,angle=0,scale=0.9]{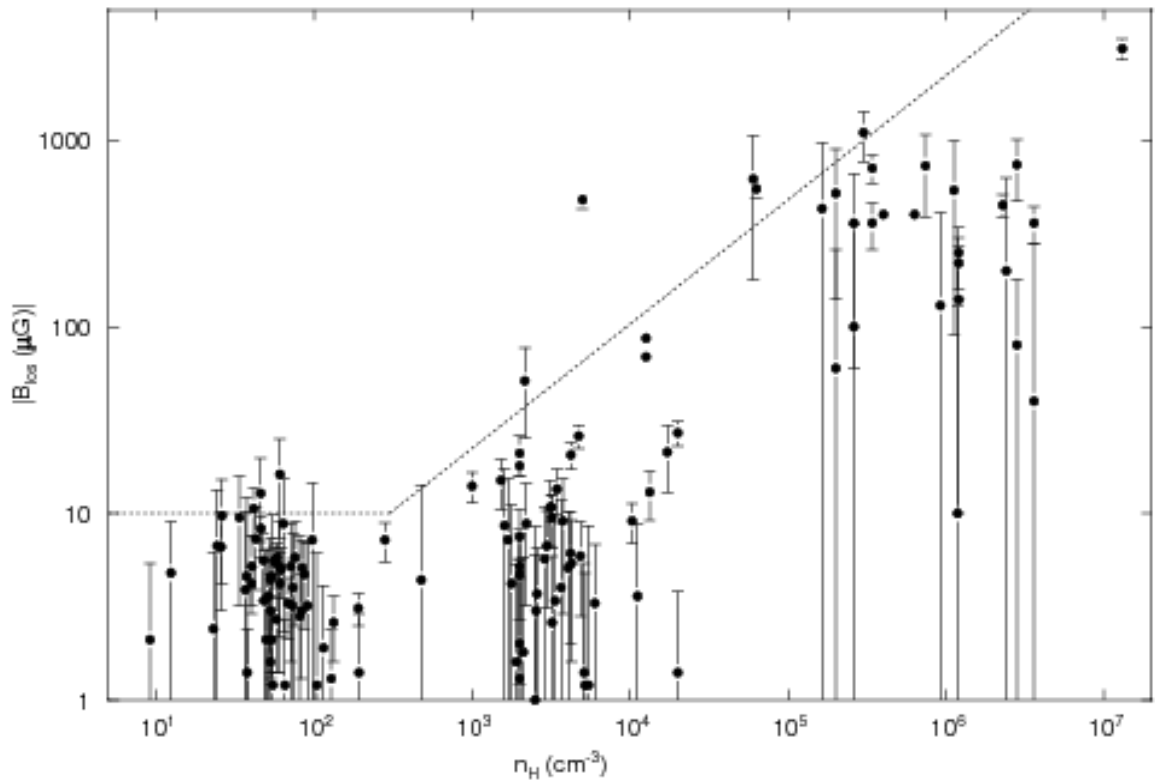}
\caption[Line-of-sight magnetic field strength versus volume density]{The set of diffuse cloud and molecular cloud Zeeman  measurements of the magnitude of the line-of-sight component $B_{LOS}$ of the magnetic vector {\bf B} and their $1\sigma$ uncertainties, plotted against $n_H = n(H~I)$ or $2n(H_2)$ for H~I and molecular clouds, respectively. Different symbols denote the nature of the cloud and source of the measurement: H~I diffuse clouds, filled circles \cite{2004ApJS..151..271H}; dark clouds, open circles \cite{2008ApJ...680..457T}; dark clouds, open squares \cite{1999ApJ...520..706C}, molecular clouds, filled squares \cite{1999ApJ...520..706C}; and molecular clouds, stars \cite{2008A&A...487..247F}. Although Zeeman measurements give the direction of the line-of-sight component as well as the magnitude, only the magnitudes are plotted. The dotted line shows the most probable maximum values for $B_{TOT}(n_H)$ determined from the plotted values of $B_{LOS}$ by the Bayesian analysis of \cite{2010ApJ...725..466C}.}
\label{Bvsn}
\end{figure}

\begin{figure}[t!]
\includegraphics[bb= 0 148 700 550,clip,scale=0.7]{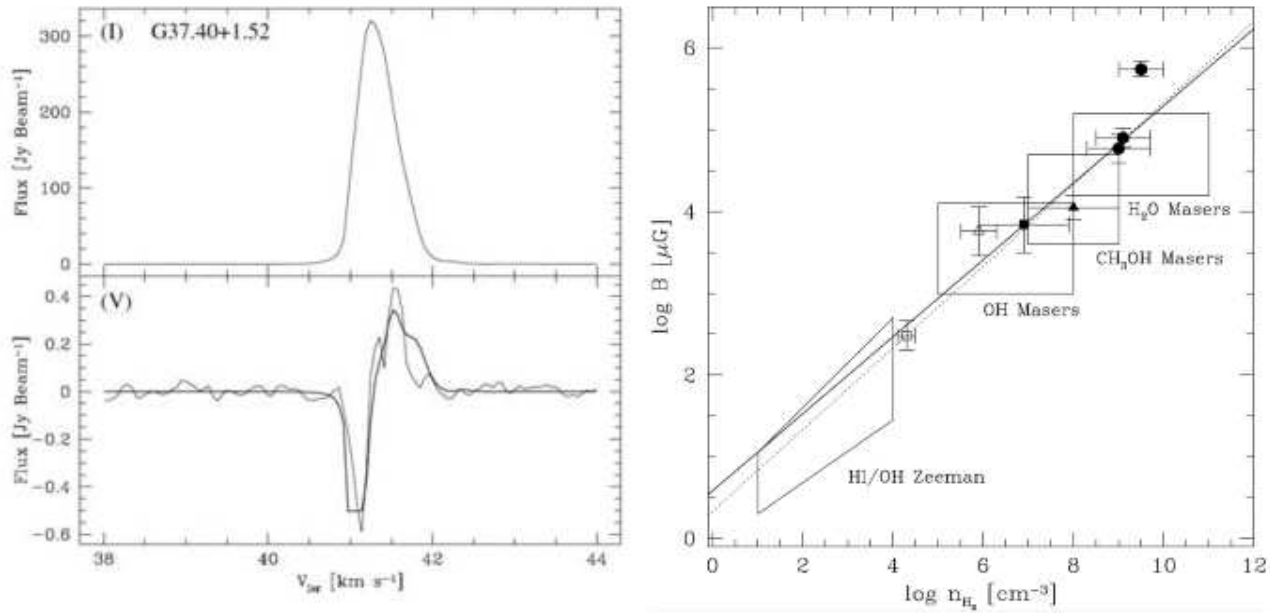} 
\caption[Maser Zeeman results]{Maser Zeeman results from \cite{vlemmings_new_2008}. Left: Example of 6.7 GHz methanol maser Zeeman observation. Inferred $B_{LOS} = 15.4 \pm 0.4$ mG. Right: Points with error bars show maser Zeeman inferred values of magnetic field strength in the massive star forming region Cepheus A. Boxes show literature values for maser and extended regions for other regions. Solid line show $B \propto n^{0.5}$ relation while dotted line shows fit to Cepheus A data.  Although the slope from this figure seems different from the 0.65 from figure \ref{Bvsn}, the maser slope does not allow for the constant value of field strength for H~I Zeeman data. When this is included, the slope for the maser gas is reasonably compatible with 0.65 given the large uncertainties in the gas densities in the maser regions.}
\label{Masers}
\end{figure}



\end{document}